\documentclass{ws-ijmpa}

\begin{document}

\markboth{Authors' Names}
{Instructions for Typing Manuscripts (Paper's Title)}

%
\catchline{}{}{}{}{}
%

\title{STRUCTURE OF LIGHT AND HEAVY PENTAQUARKS
\footnote{
Plenary talk at the 10th International Symposium on {\it Meson-Nucleon 
Physics and the Structure of the Nucleon}, Beijing, August 29 - September 4, 
2004.}
}

\author{Fl. STANCU
}

\address{Physics Department, University of Liege , \\
Sart Tilman, B-4000 Liege 1, Belgium
}

\maketitle


\begin{abstract}
Light and heavy pentaquarks are described within a constituent 
quark model based on a spin-flavor hyperfine interaction.  
In this model the lowest state acquires positive parity.
The masses of the light antidecuplet members are calculated
dynamically using a variational method.
It is shown that the octet and
antidecuplet states with the same quantum numbers mix ideally 
due to SU(3)$_F$ breaking. Masses of the charmed antisextet 
pentaquarks are predicted within the same model.
\end{abstract}

\section{Introduction}	

The pentaquarks have been discussed in the literature since more
than 30 years. Light and heavy pentaquarks have alternatively 
been predicted and searched for. The present wave of interest came
with the observation of a narrow resonance, called $\Theta^+$, of mass
$M \simeq 1540\pm10$ MeV and width $\Gamma < 25$ MeV,
by the LEPS collaboration at the end of 2002
\cite{NAKANO1}. This was interpreted as a light pentaquark of content 
$uudd \bar s$ following the 1997 predictions of Diakonov, Petrov and
Polyakov \cite{DPP} (for some other pioneering work see \cite{DIAKONOV}).
The LEPS experiment has been followed by the observation of another narrow
resonance, 
called $\Xi^{--}$, M $ \simeq$ 1862 MeV, $\Gamma \simeq$ 18 MeV,
by the NA49 collaboration \cite{NA49}, interpreted 
as a pentaquark of content $ddss \bar u$ and supposed to be also
a member of the antidecuplet of $\Theta^+$.
In March 2004 a narrow heavy resonance, 
$M \simeq$ 3100 MeV, $\Gamma \simeq$ 12 MeV, has been
observed by the H1 Collaboration \cite{AKTAS} and was interpreted as 
a heavy pentaquark of content $uudd \bar c$. The pentaquark   $\Theta^+$
needs more solid confirmation (see Ref. \cite{NAKANO2}) while the
other observations remain much more controversial. (For a historical
note on pentaquarks see e. g. \cite{FS3})

\section{Present Approaches}
Presently there is a large variety of approaches to pentaquarks:
the chiral soliton or the Skyrme model, 
the constituent quark model, the instanton model,
QCD sum rules, lattice 
calculations, etc. The main issues are the determination of 
spin and parity, the decay width, the splitting between 
the isomultiplets belonging to the same irreducible representation
of SU(3)$_F$ and the effect of representations mixing on the 
masses and widths of pentaquarks.

\section{Constituent Quark Models}
Here I shall refer to pentaqurk studies in constituent quark 
models only.
The constituent quark models describe a large number of observables 
in baryon spectroscopy (masses, formfactors, decay widths, etc) 
and it therefore seems quite natural to look at its 
predictions for exotics. The most common constituent 
quark models have either a spin-color (CS) or
a spin-flavor (FS) hyperfine interaction. One can also have
a superposition of CS and FS interactions in the so-called
hybrid models (see below).

\section{Why the FS Model ?}
The present talk focuses mainly on a dynamical study of 
the light pentaquark antidecuplet and the charmed antisextet
based on the FS interaction. Details of this study can be found 
in Refs. \cite{FS1,SR,FS2}. The calculations are based on the
model Hamiltonian of Ref.\cite{GPP}, which represents a realistic
form of the more schematic model of Ref. \cite{GR}.
Chronologically the model Hamiltonian \cite{GPP} has been first applied to the 
description of heavy pentaquarks of both negative \cite{GENOVESE} 
and positive parity \cite{FS1}. It turned out that the positive
parity pentaquarks were much lighter (few hundreds of MeV) 
than the negative parity ones at a fixed $q^4 \bar Q$ content, 
where $Q = c$ or $b$. The parity can be found by looking at
the $q^4$ subsystem. 
The lowest negative parity state with spin $S = 1/2$ comes from
a $q^4$ subsystem which
has the structure $|[4]_O [211]_C [211]_{OC};[211]_F [22]_S [31]_{FS} \rangle$. 
This means that there is no orbital excitation and the parity of the pentaquark
is the same as that of the antiquark.  If one quark is excited to the p-shell
which implies a positive parity for the pentaquark, the Pauli
principle allows the $q^4$ subsystem to have the structure
$|[31]_O [211]_C [1111]_{OC};[22]_F [22]_S [4]_{FS} \rangle$
in its lowest state. Although this state contains one unit of
orbital excitation the attraction brought by the FS interaction
when $q^4$ is in its most symmetric state $[4]_{FS}$ is so
strong that it overcomes the excess of kinetic energy and generates
a positive parity state much below the negative parity one. 

Besides giving a good description of non-strange and strange 
baryon spectra, by bringing the Roper resonance below the first negative 
parity state and describing well other baryon properties 
(see~  e. g. Ref. \cite{RISKA}), 
the flavor-spin interaction has support 
from QCD lattice calculations on baryons which suggests that
the observed level order of positive and negative parity is an
effect of the lightness of the $u$ and $d$ quark masses. Moreover
the flavor-spin is a symmetry consistent with the large $N_c$ limit 
of QCD.
 
In CS models the lowest positive parity state for $\Theta^+$, with one  
unit of orbital excitation has  
symmetry $[31]_{CS}$ in the relevant degrees of freedom. 
Even for this symmetry, which is the lowest allowed one, the hyperfine 
interaction
is not strong enough to overcome the excess of kinetic energy, 
so that the lowest state 
acquires negative parity in realistic models \cite{TS}.
For various reasons,
hybrid models also favor negative parity \cite{ZHANG}.

\section{The light pentaquark antidecuplet}

In SU(3)$_F$
a pentaquark state, described as a $q^4 \bar q$ system can be obtained 
from the direct product
\begin{eqnarray}\label{REPR}
3_F \times 3_F \times 3_F \times 3_F \times {\bar 3}_F
 & = & 
3 (1_F) + 8 (8_F) + 3 (27_F) + 4 (10_F) \nonumber \\
&+& 2 ({\overline {10}}_F)
 + 3(27_F) + 2 (35_F)
\nonumber 
\end{eqnarray}
which shows that the antidecuplet  ${\overline {10}}_F$  is one of the possible
multiplets. The SU(3)$_F$ breaking induces representation
mixing. One expects an important mixing between octet members and antidecuplet
members with the same quantum numbers. This holds for $I = 1/2$ and 1.
The results for the mass spectra of ${\overline {10}}_F$ and $8_F$
pentaquarks and their mixing based on the model of Ref. \cite{GPP}
are shown in Fig. 1.  Here,
as in any other model including the chiral soliton, 
one cannot determine the absolute mass of 
$\Theta^+$. This mass has been fitted to the presently accepted 
experimental value of 1540 MeV. Reasons to accommodate such  
a value are given in Ref. \cite{SR}. The pure ${\overline {10}}_F$
spectrum, Fig. 1a, can approximately be described by the linear mass
formula $M = 1829 - 145 Y$ where $Y$ is the hypercharge. This result 
is compared to the antidecuplet spectrum of Ref. \cite{DPP},
~Fig. 1c, where
$M = 1829 - 180 Y$, which implies larger spacing than in the FS model. 
Presently the level spacing in the chiral
soliton model is estimated to be smaller \cite{ELLIS}, thus closer 
to the FS model result. 

As a consequence of the SU(3)$_F$ breaking the representations
${\overline {10}}_F$ and $8_F$ mix and accordingly 
the physical states are defined as
\begin{eqnarray}\label{PHYSN}
|N^*\rangle  = |N_{8} \rangle \cos \theta_N
 - |N_{\overline {10}}\rangle \sin \theta_N,\nonumber \\
|N_5\rangle =  |N_{8}\rangle \sin \theta_N 
+ |N_{\overline {10}}\rangle \cos \theta_N,
\end{eqnarray} 
for $N$ and similarly for $\Sigma$.
Fig. 1b shows the masses of the physical states  $|N_5\rangle$
obtained from this mixing 
${\overline {10}}_F$ and $8_F$ \cite{FS2}. 
The mixing angles $\theta_N$
and  $\theta_{\Sigma}$ are calculated dynamically in this approach,
by using the coupling matrix element of ${\overline {10}}_F$ and $8_F$. 
This has combined contributions from all the parts of the Hamiltonian
which break SU(3)$_F$ symmetry: the free mass term, the kinetic energy and the
hyperfine interaction, see Ref. \cite{FS2}. Interestingly, the combined 
contribution 
leads in practice to an ideal mixing. One obtains $\theta_N \simeq 35.34^0 $,
as compared to the ideal mixing angle $\theta_N^{id} = 35.26^0$  and
$\theta_{\Sigma} \simeq - 35.48^0 $ as compared to 
$\theta_{\Sigma}^{id} = - 35.26^0$. 

By the definition (\ref{PHYSN}),
~ $|N^*\rangle$ is the ``mainly octet'' state and 
$|N_5\rangle$ is the ``mainly  
antidecuplet'' state.
From the calculated mixing angle it follows
that the former is approximately 67 \% octet and the latter 
67 \% antidecuplet. The above  mixing angle implies 
that $\cos \theta_N
\simeq \sqrt{2/3}$ and  $\sin \theta_N  \simeq \sqrt{1/3}$. 
Then, for example, for positive charge pentaquarks with $I$ = 1/2
one has
\begin{eqnarray}
|N^*\rangle & \simeq &\frac{1}{2}~(ud - du)(ud - du) \bar d,\nonumber \\
|N_5\rangle & \simeq &\frac{1}{2 \sqrt{2}}~[(ud - du)(us - su) + (us - su)(ud - du)] 
\bar s,
\end{eqnarray} 
i. e. the ``mainly octet" has no strangeness and the 
``mainly antidecuplet'' state contains the whole available 
amount of (hidden) strangeness.
 
The ``mainly octet'' nucleon state $|N^*\rangle$  acquires the
mass M(N$^*$) $\simeq $ = 1451 MeV and for the ``mainly octet'' $\Sigma$ state
one obtains M($\Sigma^*$) $\simeq $ = 2046 MeV. In the nucleon case
this implies that there are two resonances in the Roper mass region:
one which is obtained in the original calculations of Ref. \cite{GPP}
as a radial excited state 
of structure $q^3$ and mass 1495 MeV and the other the pentaquark state 
at 1451 MeV. A mixing of  $q^3$ and $q^4 \bar q$ could be a better 
description of the reality. 
There is some experimental evidence \cite{MORSCH}
that two resonances instead of one
can consistently describe the $\pi - N$ and the $\alpha - p$ scattering
in the Roper resonance region 1430 - 1470 MeV.  

\begin{figure}\label{figure1}
\centerline{\psfig{file=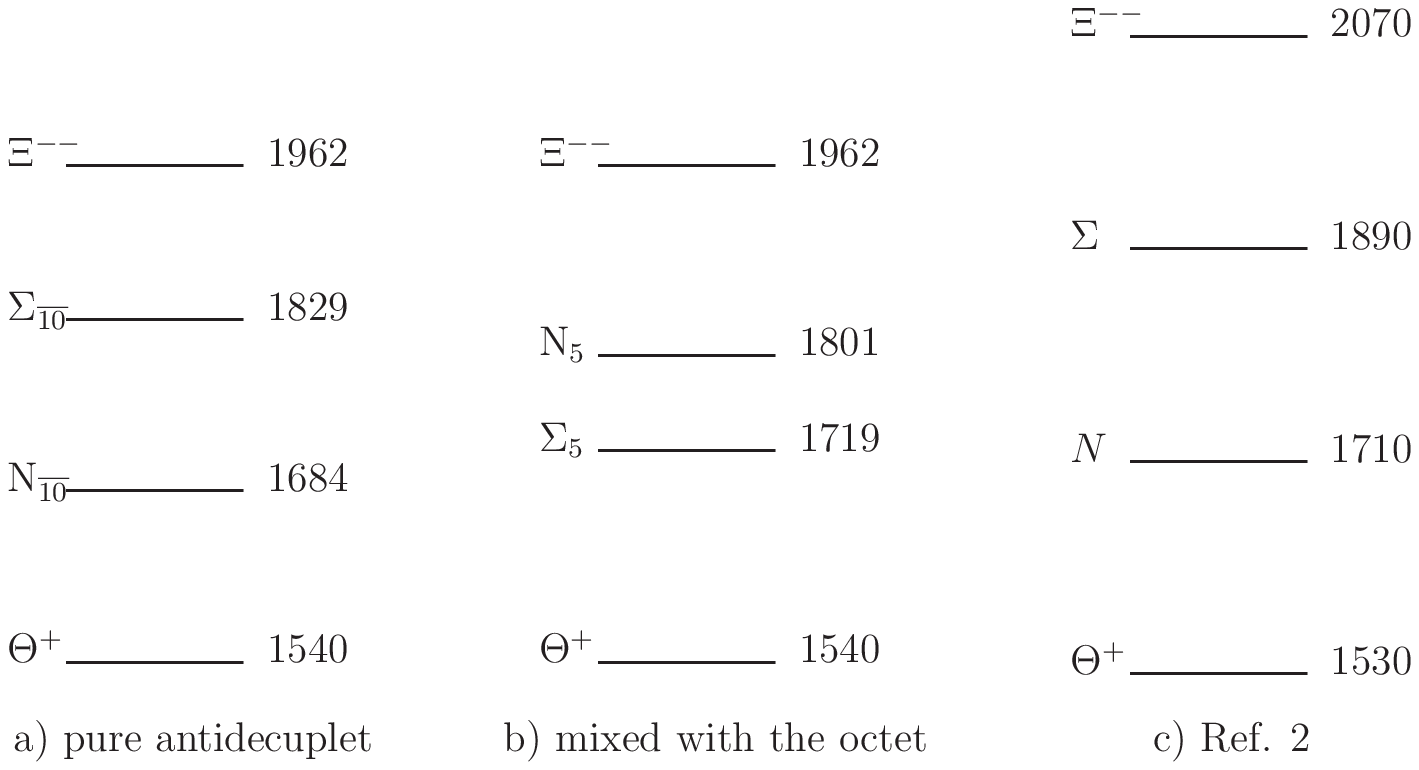,width=11cm}}
\vspace*{8pt}
\caption{The pentaquark antidecuplet masses (MeV) in the FS model:
(a) pure antidecuplet and (b)
after mixing with the pentaquark octet, as compared with
the predictions (c) of
the chiral soliton model Ref. \protect\refcite{DPP}.}
\end{figure}


\section{The charmed antisextet}

In most models which accommodate $\Theta^+$ and its antidecuplet partners, 
heavy pentaquarks $q^4 \bar Q$ can be accommodated
as well, being in principle more stable against strong decays
than the light pentaquarks \cite{FS4}. In an 
SU(4) classification, including charm, it has been shown \cite{MA} that the 
light antidecuplet containing  $\Theta^+$ and a charm antisextet
with $\bar s$ replaced by $\bar c$ belong to the same
SU(4) irreducible representation of dimension 60.

\begin{figure}\label{figure2}
\centerline{\psfig{file=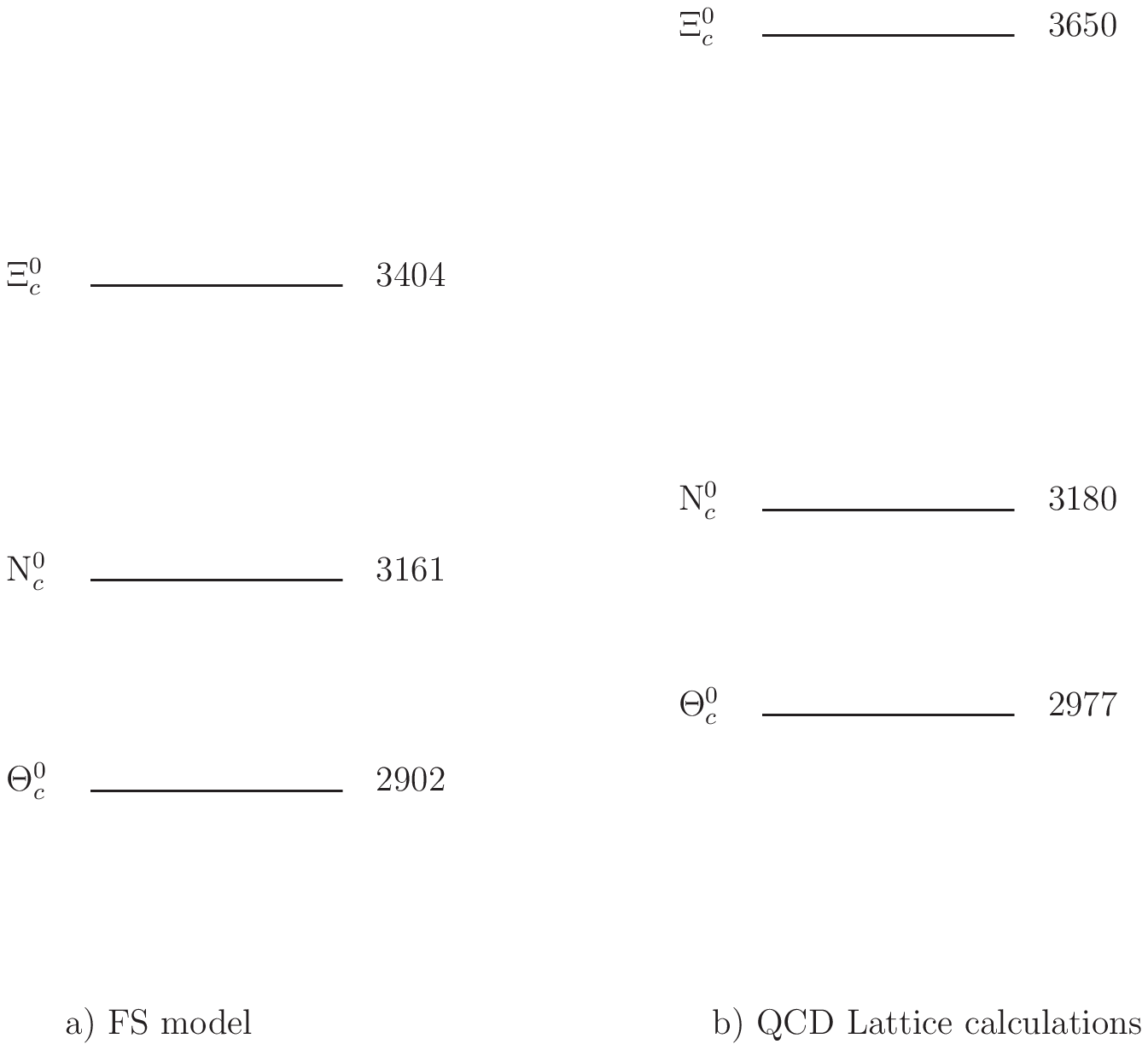,width=9cm}}
\vspace*{8pt}
\caption{The charmed pentaquark antisextet masses (MeV):  
 (a) The FS model results of Ref. \protect\refcite{FS1}
and (b) QCD lattice calculations of
Ref. \protect\refcite{CHIU}.}
\end{figure}

The masses of the charmed antisextet calculated in the FS model 
\cite{FS1} are presented in Fig. 2. They are compared to the 
only QCD lattice calculations \cite{CHIU} which predict positive
parity for the lowest states. Although the two calculations 
have an entirely different basis the masses of $\Theta^0_c$
of content $uudd \bar c ~(I = 0)$ and of $N^0_c$ of content 
$ uuds \bar c ~(I = 1/2)$
are quite similar. Only for $\Xi^0_c$ of content $ uuss \bar c ~(I = 1)$
a mass difference of 250 MeV appears. 
However, by taking into account the lattice calculations numerical
uncertainty of approximately $\pm$ 100 MeV, the two results look
closer to each other.
The experimental observation
of charmed pentaquarks is contradictory so far. While the 
H1 collaboration \cite{AKTAS} brought evidence for a narrow resonance
at  about 3100 MeV, there is null evidence from ZEUS  
or the CDF collaborations \cite{WWW2}.

For an orientation, in connection to future experiments,
it is interesting to calculate excited charmed pentaquarks
$\Theta^{0*}_c$. In the FS model used above the first excited state
having $I = 1$ and $S = 1/2$ appears 200 Mev above $\Theta^{0}_c$
which has $I = 0$. This supports the large spacing result
obtained approximately in Ref. \cite{MALTMAN} for the FS model.

\section{Conclusions}

The recent research activity on pentaquarks 
could bring a substantial progress in 
understanding the baryon structure. It could help to shed a new light on
some resonances difficult to understand, as e. g. the Roper or the 
$\Lambda(1405)$ baryon. The wave functions of such baryons
may have substantial higher Fock components.
Furthermore, in light hadrons it is important to understand the role  
of the spontaneous breaking of chiral symmetry, 
the basic feature of the chiral soliton model \cite{DPP}
which motivated this new wave of interest in pentaquarks.
It is thought that the FS model used here has its origin
in the spontaneous breaking of chiral symmetry as well \cite{GR}.
An important common feature of the two models is that
they both lead to positive parity for the light antidecuplet. 
If ever measured, the parity of $\Theta^+$ could disentangle
between various models. Also a clear theoretical understanding 
of the production mechanism of light and heavy pentaquarks 
is necessary.  
The experiments, which see or not see the pentaquarks, need a
clear explanation. 

\section*{Acknowledgments}

I am most grateful to the organizers of the MENU04 Conference 
for their kind invitation and the excellent organization of
this meeting.

\end{document}